# BERNOULLI EFFECT AT FREE REVOLVING FLOW BRAKING

V.A.Budarin

*National Polytechnical University, Odesa, Ukraine*

In this research paper, analytical equations for the calculation of radial flow due to the braking on the flat surface plane of revolving flow were obtained. The calculation method is based on the use of pressure force balance equation, viscid friction and inertia. Three motion equations for incompressible, polytropic ($n \neq 1$) and isothermal flow were obtained; the type of a cumulative curve for the incompressible flow has been shown. Possibility of the use of radial flow for the compression of water vapour cloud has been shown.

Keywords: *radial flow, revolving flow, cloud compression.*

## I. Introduction

A scientific problem about the vortex filament braking on a boundless plane and the initiation of wall flow was examined in many works [1,2]. On this condition, it was assumed that the axis of vortex filament is perpendicular to the plane, and all its parameters are known. While approaching the sectional view of vortex filament to the braking surface, its rotation speed diminishes under the impact of viscid friction forces. It results in reduction of centrifugal force and change of force balance, which are characteristic of sections that are distant from a surface where an influence of viscid friction forces can be negligible.

Such a flow is known in meteorology as it appears at the contact of lower butt end of a tornado with a surface and is called sole [3]. In the power engineering, an analogical flow appears in vortex chambers and is called butt-end effect. Different manifestations of this effect have identical physical cause. While braking the revolving flow nearby a surface, the wall radial flow directed from periphery to braking area appears there. The findings show that the rotation speed of vortex filament on a wall is equal to the zero that is compliant with hypothesis of adhesion and allows neglecting the influence of tangential velocity component outside the braking area. Such assumptions allow considering a wall flow as two-dimensional and axisymmetrical [1].

A sum about the calculation of a wall flow under the given terms can be done by means of integration of Navier-Stokes' equation. This solution, like all other analogical exact solutions are in strict compliance with findings at Reynolds' small numbers only. For this sum of *Re* $_{max}$ = 5, that is obviously associated with an approximate character of Navier-Stokes' differential equation [4].

## 2. Design model and derivation of motion equations

The calculation of wall flow that arises from braking the vortex tube flow of finite sizes is examined in the paper. The aim of calculation is to find the pressure distribution along the radius of wall flow and estimate the pressure force directed perpendicularly to the surface. The method of calculation is based on the use of equilibrium equation of small flow element being under the power of permanent pressure forces, viscid friction and inertia.

Let us do the following sum supposing that the wall flow is discontinuous, i.e. inside the flow, there exists the closed path with a radius of $r_2$, where there is a jump of radial speed there, and pressure is considered permanent (fig.I). The presence of such a path is caused by geometry of flow motion, which, while approaching the axis, turns upwards and convolves in a spiral.

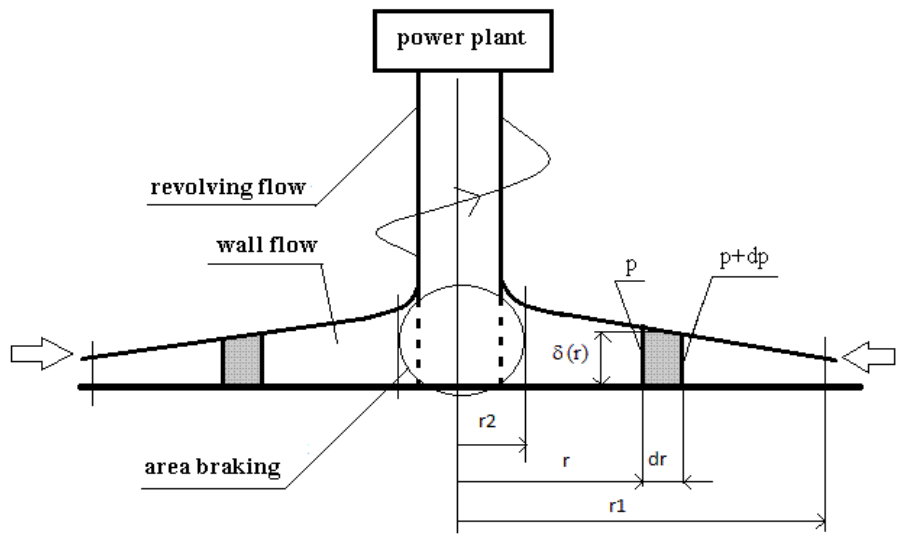

Figs.I  Design model of flat axisymmetrical flow on a round plate.

The analysis of the findings obtained regarding such a flow showed that the tangential speed component of wall flow subsides quickly as far as the increase in radius is concerned. And the speed distribution of the radial flow is close to potential, i.e. $u \cdot r = const.$

The given hypothesis allow to work out the following equation of forces balance that have an impact on the elementary ring in width $dr$ and in hight $r$ (figs. 1)

$$\delta P_{pr} + \delta P_{fr} + \delta P_{in} = 0 \tag{1}$$

The use of design model results in the following expressions for elementary forces.

$$\delta P_{pr} = 2\pi\delta \left[ p \cdot r - (p+dp)(r+dr) \right]$$

$$\delta P_{in} = dm \cdot a_r = \pi\rho\delta \left[ (r+dr)^2 - r^2 \right] \left( u \frac{\partial u}{\partial r} \right)$$

Using the Newton's law for a viscid friction

$$\delta P_{fr} = \mu \cdot df \cdot gradu = \mu\pi \left[ (r+dr)^2 - r^2 \right] \frac{\partial u}{\partial \delta}$$

— elementary pressure forces, inertia and viscid friction accordingly; $\mu$ is dynamic viscidity; $\delta(r)$ is a thickness of wall flow; $u$ is everage velocity; $a_r$ is a convection acceleration.

After a substitution of forces balance in the equation, we will get the folowing differential motion equation of the wall flow along the radius of the plate:

$$\frac{\partial p}{\partial r} + \frac{p}{r} = \rho \left( u \frac{\partial u}{\partial r} + \frac{v}{\delta} \frac{\partial u}{\partial \delta} \right) \tag{1}$$

For the exact sum of this equation, it is necessary to know the law of variation of flow thickness along a radius. Substitution of the known equation for a plate at its parallel flowing around

$$\delta(r) = 0{,}5 \cdot \left[ \frac{v \cdot (r_1 - r)}{u} \right]^{0,5}$$

does not give a satisfactory effect, as it results in reduction of kinematics viscidity [5]. This fact indicates that there exists the substantial differences of the plane-parallel flowing around of plate from radial. In addition, the adjustment for viscidity influence only does not reflect influence of roughness and surface profiling.

Let us consider another way of calculation of friction force that uses a complex parameter – a coefficient of resistance at the radial flow of plane $C_в$. Then

$$\delta P_{fr} = C_r \cdot df \cdot \frac{\rho u^2}{2} = C_r \pi \left[ (r + dr)^2 - r^2 \right] \frac{\rho u^2}{2}$$

Differential equation for the incompressible liquid will go over:

$$\frac{dp}{dr} + \frac{p}{r} = \rho \left( u \frac{du}{dr} + \frac{C_r}{\delta} \frac{u^2}{2} \right) \qquad (2)$$

This equation has an exact solution as we know the velocity variation along a radius, and dependence of coefficient of $C_в$ on the velocity is fundamentally known or approximately can be accepted as permanent. The thickness of wall flow can be found from equation of mass balance.

$$\delta = \frac{m}{2\pi \rho u r}$$

Because $-u \cdot r = const$ the mass expense of $m$ and density of $p$ do not change in the incompressible flow, then $\delta \approx const$.

Equations (2) and (3) can be modified in order to describe the flow of ideal gas for a polytropic ($n \neq 1$) and isothermal flow ($n = 1$).

Because $\rho = \rho_2 \left( \dfrac{p}{p_2} \right)^{1/n}$

$$\frac{dp}{dr} + \frac{p}{r} = p^{1/n} \cdot \frac{\rho_2}{p^{1/n}_2} \left( u \frac{du}{dr} + \frac{C_r}{\delta} \frac{u^2}{2} \right) \quad (4)$$

For an isothermal flow $\rho = \dfrac{p}{RT}$

$$\frac{dp}{dr} + \frac{p}{r} = p^{1/n} \cdot \frac{\rho_2}{p^{1/n}_2} \left( u \frac{du}{dr} + \frac{C_r}{\delta} \frac{u^2}{2} \right) \quad (5)$$

Equations (3), (4) and (5) have exact solutions. For example, equation (3) over can be gone over

$$\frac{dp}{dr} + \frac{p}{r} = A \cdot \left[ \frac{B}{r^2} - \frac{1}{r^3} \right]$$

$[\, u = \dfrac{u_2 r_2}{r}, \; A = \rho \cdot u_2^{\,2} \cdot r_2^{\,2}, \; B = \dfrac{C_r}{2\delta} \cdot ]$

### 3. Discussion

3.1. Equation (2) cannot be obtained by simplification of Navier-Stokes' equation that is being discussed in the paper [4].

3.2. Presently, there are no reliable data about the variation of thickness of wall flow along a radius $\delta(r)$, as well as data about the size of coefficient of resistance at the radial flow of $C_2$. Finding these values requires that the special tests be carried out.

3.3. In fig.2, a general integral of differential equation for an incompressible liquid, type of integral curve, and quantitative estimation of Bernoulli effect, calculated in the package of Maple is shown.

```
> ode:=diff(p(r),r)+p(r)/r=A*(B*r^(-2)-r^(-3));
```
$$ode := \frac{d}{dr} p(r) + \frac{p(r)}{r} = A \left( \frac{B}{r^2} - \frac{1}{r^3} \right)$$

```
> dsolve(ode);
>
```
$$p(r) = \frac{-\frac{A}{r} + A B \ln(r) + \_C1}{r}$$

```
> A = 0.3, B = 0.02, C1 = -100
```
$$A = 0.3, B = 0.02, C1 = -100$$

```
> plot
```
$$plot\left( \left( \frac{\frac{-0.3}{r} + 0.006 \cdot \ln(r) - 100}{r} \right), r = 0.002 .. 0.3 \right);$$

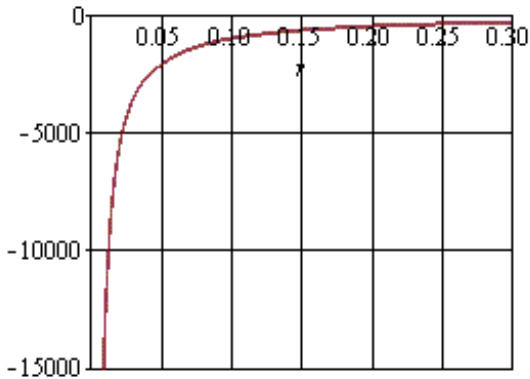

$$L := Int\left( 6.28 \cdot \left( \frac{-0.3}{r} + 0.006 \cdot \ln(r) - 100 \right), r = 0.2 .. 0.02 \right)$$
$$= int\left( 6.28 \cdot \left( \frac{-0.3}{r} + 0.006 \cdot \ln(r) - 100 \right), r = 0.2 .. 0.02 \right);$$

$$L := \int_{,20}^{,02} \left( -\frac{1,88}{r} + ,04 \ln(r) - 628,00 \right) dr = 117,39$$

$$> q := \frac{L}{3.14 \cdot (0.2^2 - 0.02^2)};$$

$$q := 8,04 \left( \int_{,20}^{,02} \left( -\frac{1,88}{r} + ,38 \ln(r) - 628,00 \right) dr \right) = 945,26$$

Fig. 2. General integral and cumulative curve for an incompressible flow. Force of pressure of $L$(N) and surface load of $q$(N/m2).

3.4. As shown in fig.3, there is variant of application of wall flow for the water vapour cloud compression. The tubular airflow had a diameter of about *1 cm* and it was created by a compressor with electric power of *300 W*.

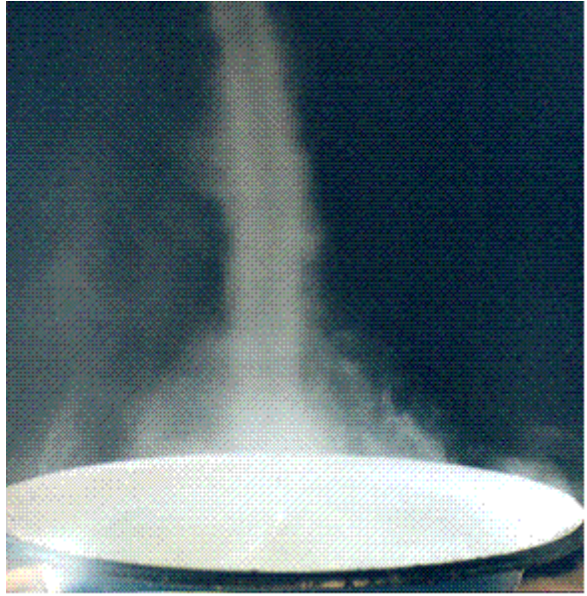

Fig.3. Photo of water steam cloud compressing under the power of radial wall flow.